\documentclass[fleqn,10pt]{wlscirep}
\usepackage[utf8]{inputenc}
\usepackage[T1]{fontenc}

\usepackage{amsmath}
\usepackage{graphicx}
\usepackage{dcolumn}
\usepackage{rotating}
\usepackage[english]{babel}

\usepackage{color}
\usepackage{adjustbox}
\usepackage{float}

\title{Chemical bonding in americium oxides: x-ray spectroscopic view}

\author[1,*]{Sergei M. Butorin}
\author{David K. Shuh}

\affil[1]{Condensed Matter Physics of Energy Materials, X-ray Photon Science, Department of Physics and Astronomy, Uppsala University, P.O. Box 516, SE-751 20 Uppsala, Sweden}
\affil[2]{Chemical Sciences Division, Lawrence Berkeley National Laboratory, MS 70A1150, One Cyclotron Road, Berkeley, CA 94720, USA}
\affil[*]{sergei.butorin@physics.uu.se}

\begin{abstract}
The electronic structure and the chemical state in Am binary oxides and Am-doped UO$_2$ were studied by means of x-ray absorption spectroscopy at shallow Am core ($4d$ and $5d$) edges. In particular, the Am $5f$ states were probed and the nature of their bonding to the oxygen states was analyzed. The interpretation of the experimental data was supported by the Anderson impurity model (AIM) calculations which took into account the full multiplet structure due to the interaction between $5f$ electrons as well as the interaction with the core hole. The sensitivity of the branching ratio of the Am $4d_{3/2}$ and $4d_{5/2}$ x-ray absorption lines to the chemical state of Am was shown using Am binary oxides as reference systems. The observed ratio for Am-doped UO$_2$ suggests that at least at low Am concentrations, americium is in the Am(III) state in the UO$_2$ lattice. To confirm the validity of the applied AIM approach, the analysis of the Am $4f$ x-ray photoelectron spectra of AmO$_2$ and Am$_2$O$_3$ was also performed which revealed a good agreement between experiment and calculations. As a whole, AmO$_2$ can be classified as the charge-transfer compound with the $5f$ occupancy ($n_f$) equal to 5.73 electrons, while Am$_2$O$_3$ is rather a Mott-Hubbard system with $n_f$=6.05.
\end{abstract}

\begin{document}

\flushbottom
\maketitle
%
%
\thispagestyle{empty}

\section*{Introduction}
The americium oxides are the important part of the nuclear fuel cycle. In the framework of the fourth generation (GEN IV) nuclear reactor development, innovative fuel cycles are currently explored. The two main goals are an efficient use of the energy resources by recycling the major actinides (An) together, such as U and Pu, and a decrease of the waste radiotoxicity by partitioning and transmutating the minor actinides, such as Am and Cm, as a part of the mixed-oxide (MOX) nuclear fuel. In this case, the studies of the incorporation of minor actinides in the lattice of (U,Pu)O$_2$ and changes in the chemical state of actinides become important. Furthermore, the assessment of the properties of MOX as the multicomponent systems requires a comprehensive knowledge of properties of each binary oxide. The americium oxides and the MOX material with Am are considered as efficient power sources for missions into deep space \cite{Wiss,Vigier,Wiss02}. That also requires detailed studies of oxide properties to help with the evaluation of their long-term performance.

From the electronic structure point of view, the character of the ground state, the strength of Coulomb interaction $U_{ff}$ between the An $5f$ electrons, the An $5f$ occupancy and degree of covalency of the An $5f$-O $2p$ bonds are important factors which affect both low-energy thermodynamic and high-energy optical properties of the system in question. X-ray methods, such as x-ray absorption spectroscopy (XAS) and x-ray photoelectron spectroscopy (XPS), are common tools in studies of electronic structure. Besides probing the chemical state of actinides in various systems, valuable information can be obtained about the oxygen/metal (O/M) ratio, (non)stoichiometry, and charge distribution, which are the parameters important for the fuel performance. However, due to high radioactivity of Am oxides, x-ray spectroscopic experiments are mostly conducted in the hard x-ray range where various containments for the samples can be used. The XAS measurements are usually performed at the Am $L_3$ edge \cite{Nishi,Prieur,Vespa,Prieur02,Prieur03,Lebreton,Prieur04,Epifano}. In this case, the Am $6d$ states are probed and the information about the $5f$ states can be obtained only indirectly. While the chemical shift of the Am $L_3$ XAS spectra is commonly used to evaluate the Am oxidation state, it was also pointed out \cite{Butorin} that the chemical shift of the spectra can be in part mimicked by a significant redistribution of the unoccupied density of states (DOS) in vicinity of the conduction band minimum. The statement was based on the high-resolution XAS data measured at the An $N_{6,7}$ edges of the An binary oxides \cite{Butorin} which also probe the An $6d$ states.

To involve the Am $5f$ states into the spectroscopic process directly, the XAS experiments at the Am $M_{4,5}$ or $N_{4,5}$ or $O_{4,5}$ are necessary. It has been shown that the sensitivity of the XAS method can be significantly improved by performing the so-called high energy resolution fluorescence detected x-ray absorption (HERFD-XAS) measurements at the An $M_{4,5}$ edges \cite{Kvashnina,Kvashnina02,Butorin02} but, in particular for Am compounds, very few attempts for such analysis were made so far \cite{Epifano,Butorin03,Butorin04}. Here, we present the results of the XAS measurements at the Am $N_{4,5}$ and $O_{4,5}$ edges of the Am oxides.

The analysis of the spectroscopic data in the framework of the Anderson impurity model (AIM) \cite{Anderson} can help to obtain information about the character of the ground state, strength of Coulomb interaction $U_{ff}$ between the Am $5f$ electrons, An $5f$ occupancy and Am $5f$-O $2p$ bonding. This is especially important in light of the discussion among the density-functional-theory (DFT) research groups about the value of $U_{ff}$ in the Am oxides. For example, different $U_{ff}$ values were claimed (varying between 4.0 eV and 7.0 eV) for the same Am oxides \cite{Wen,Suzuki,C_Suzuki,Lu,Pegg,Noutack,Chen,Moree} based on the results of DFT+$U$ calculations. Besides the AIM interpretation of the Am $N_{4,5}$ and $O_{4,5}$ XAS data of the Am oxides, we also analyzed the Am 4f XPS spectra of Am$_2$O$_3$ and AmO$_2$ within the AIM framework because such a joint analysis puts tighter restrictions on the possible values of the model parameters.

\section*{Results and Discussion}
\begin{figure}
\centering
\includegraphics[width=0.7\columnwidth]{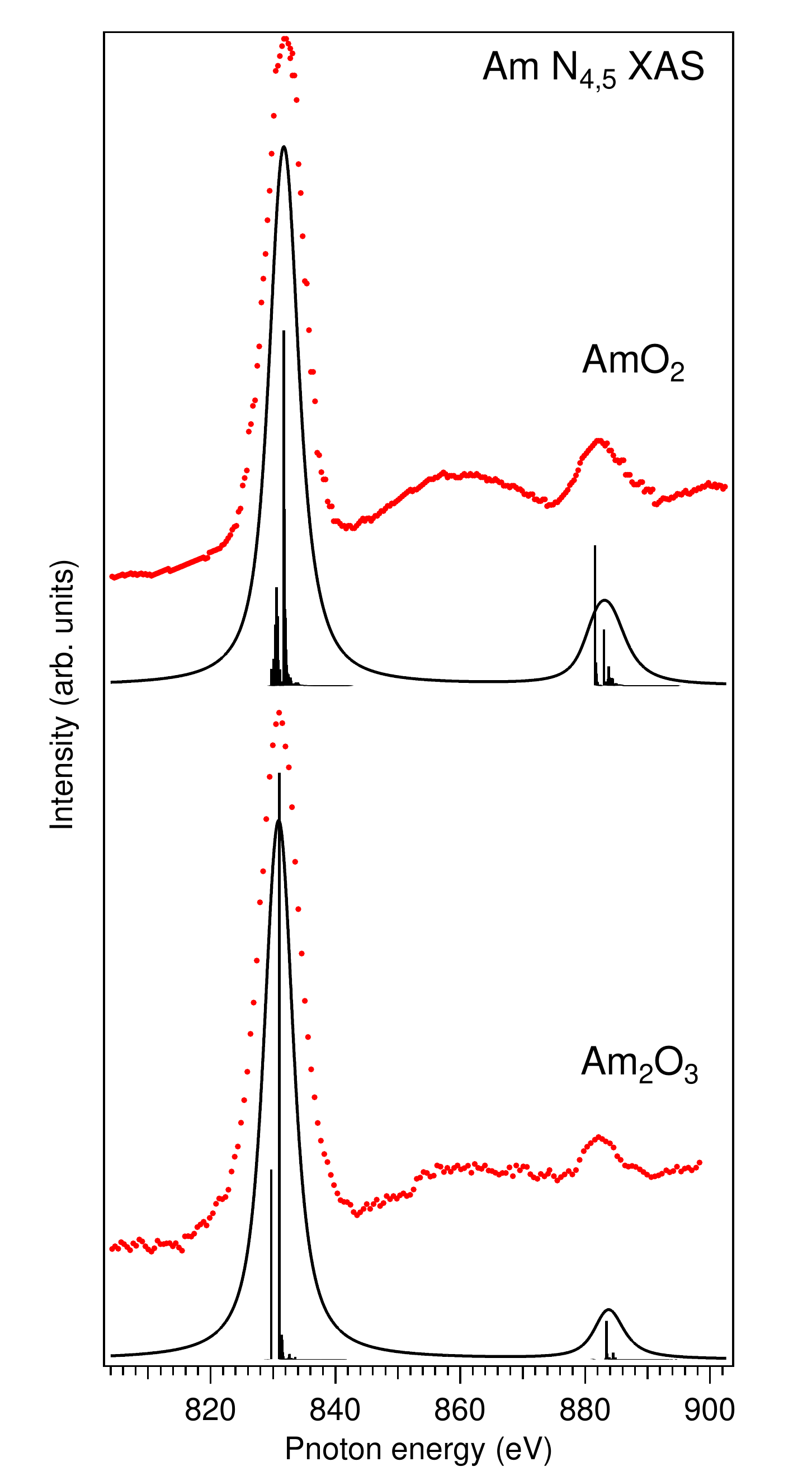}
\caption{Am $N_{4,5}$ XAS spectra of Am oxides (red markers) compared with results of AIM calculations (solid black lines and poles).}
\label{AmN45_XAS}
\end{figure}

Fig.~\ref{AmN45_XAS} displays the measured Am $N_{4,5}$ XAS spectra of Am$_2$O$_3$ and AmO$_2$. The spectra contain two main lines: $N_{5}$ ($4d_{5/2}\rightarrow5f_{7/2,5/2}$ transitions) at $\sim$831.0 eV for Am$_2$O$_3$ and at $\sim$831.8 eV for AmO$_2$, and $N_{4}$ ($4d_{3/2}\rightarrow5f_{5/2}$ transitions) at $\sim$882.0 eV for Am$_2$O$_3$ and at $\sim$882.8 eV for AmO$_2$. The intensity appearing in between the $N_{5}$ and $N_{4}$ lines represent transitions to the $7p$ states of americium. The higher intensity of the latter transitions for AmO$_2$ can be explained by the lower electron occupancy as a result of the higher oxidation state of Am. The Am $N_{4,5}$ XAS spectrum of AmO$_2$ reveals the chemical shift of $\sim$0.8 eV to the high energy side with respect to that of Am$_2$O$_3$, thus clearly indicating the difference in the oxidation state between the two samples. The value of the chemical shift is similar to that observed between the Am 4d XPS spectra \cite{Teterin} of Am$_2$O$_3$ and AmO$_2$.

\begin{figure}
\centering
\includegraphics[width=0.8\columnwidth]{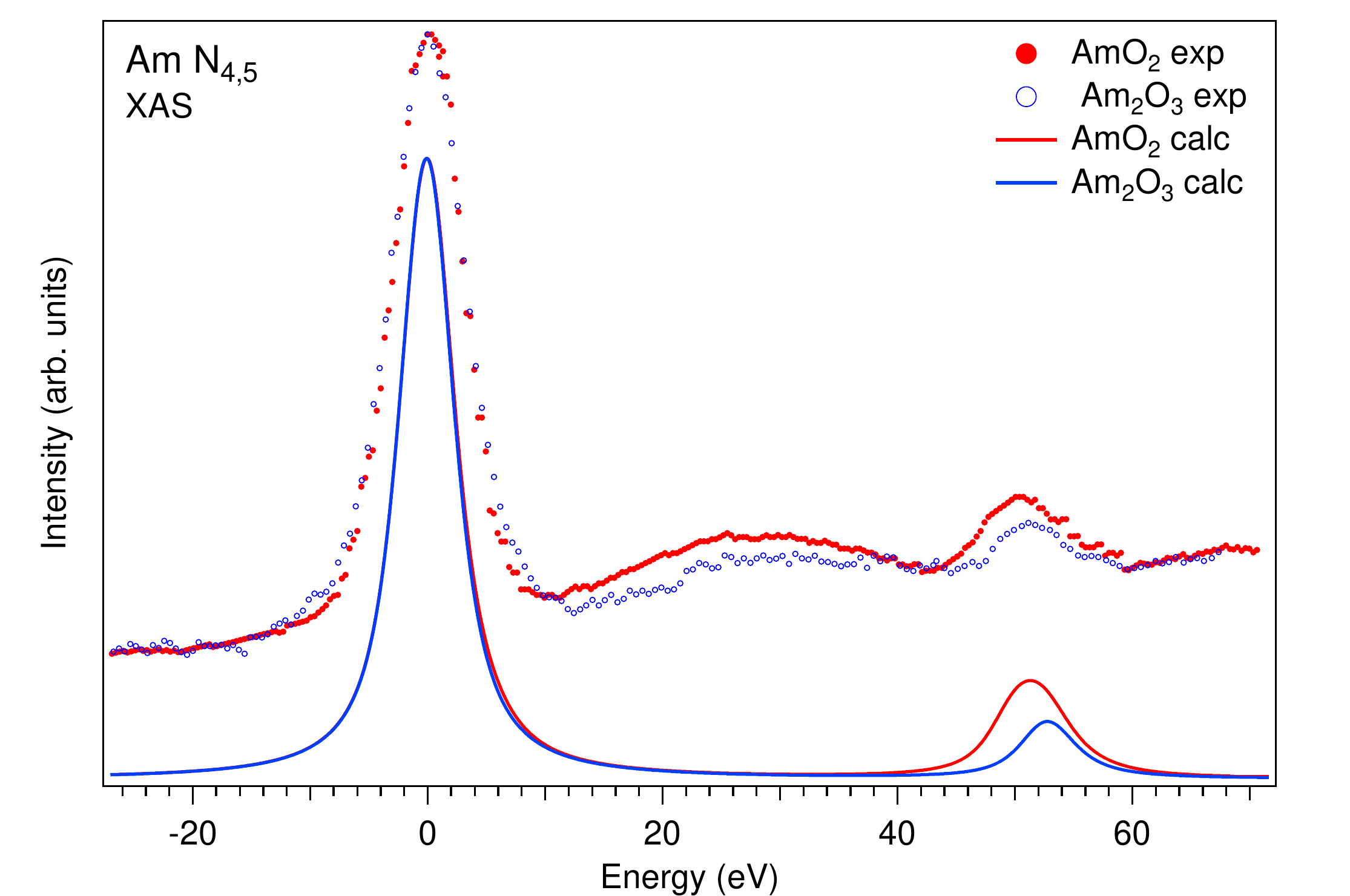}
\caption{Measured and calculated Am $N_{4,5}$ XAS spectra of Am oxides aligned to the energy of $N_{5}$ maximum (set to zero eV).}
\label{AmN45_compare}
\end{figure}

Another distinct difference between the Am $N_{4,5}$ XAS spectra of Am$_2$O$_3$ and AmO$_2$ is the intensity ratio between the $N_{4}$ and $N_{5}$ lines. This is better illustrated in Fig.~\ref{AmN45_compare}, where the Am $N_{4,5}$ XAS spectra of Am$_2$O$_3$ and AmO$_2$ are displayed on top of each other by aligning the Am $N_{5}$ maxima of Am$_2$O$_3$ and AmO$_2$. It was argued \cite{Moore,Moore02,Butorin08} that the branching ratio of the $N_{5}$ and $N_{4}$ lines, defined as $I_{5/2}/(I_{5/2}+I_{3/2})$, where $I$ is the integrated intensity of the line, is a characteristic of the An oxidation state and $5f$ occupancy/count $n_f$. A gradual decrease in the relative $N_{4}$ intensity and a corresponding increase in the branching ratio were demonstrated on going from the $n_f$ = 1 system to the $n_f$ = 6 system with reference to the nominal oxidation state/5f count. Indeed, one can see in Figs.~\ref{AmN45_XAS} and \ref{AmN45_compare} that the relative $N_{4}$ intensity is lower in the spectrum of Am$_2$O$_3$ as compared to that of AmO$_2$, thus indicating the Am(III) system versus Am(IV) one.

\begin{figure}
\centering
\includegraphics[width=0.8\columnwidth]{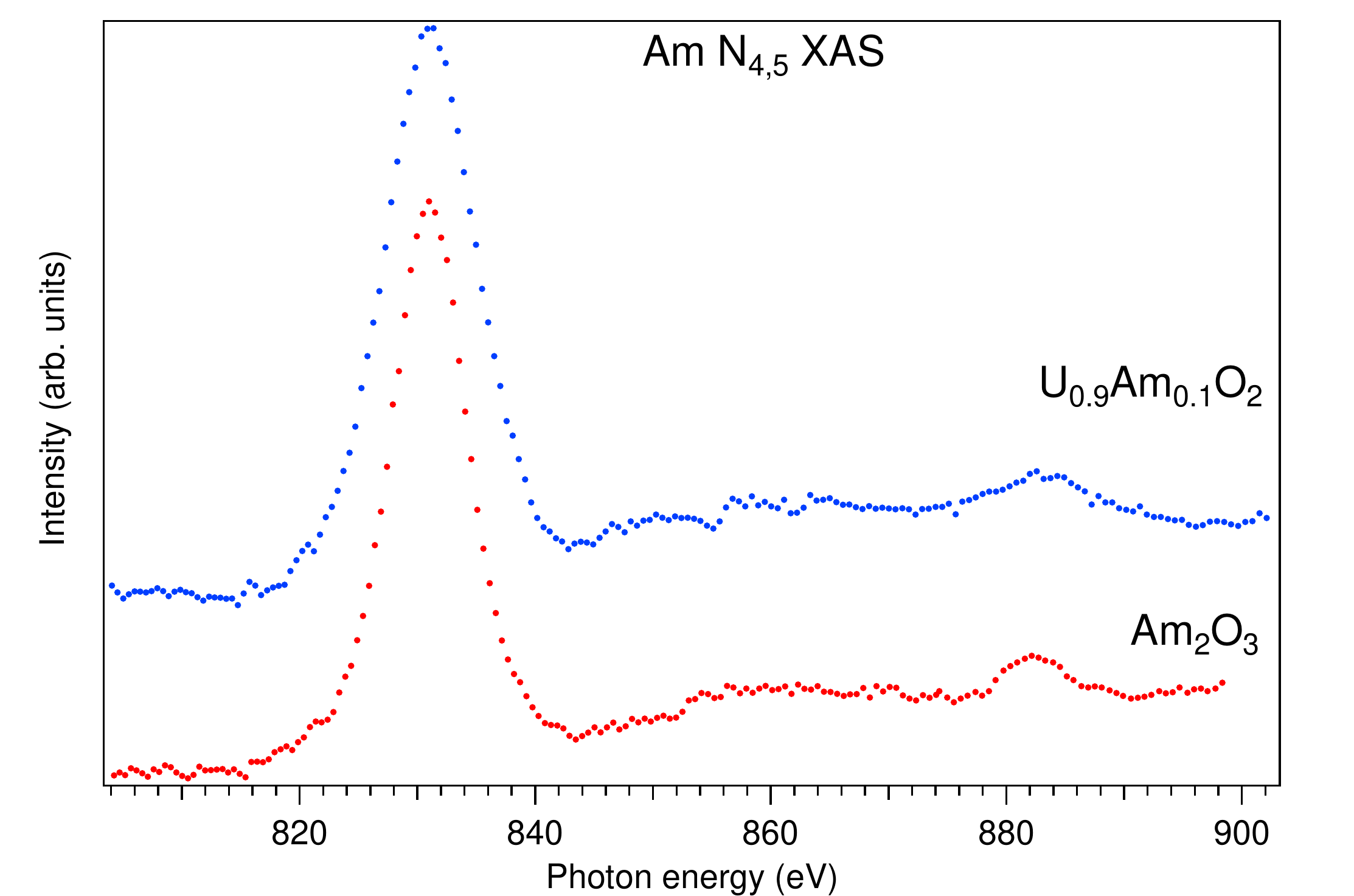}
\caption{Am $N_{4,5}$ XAS spectra of Am$_2$O$_3$ and U$_{0.9}$Am$_{0.1}$O$_2$.}
\label{AmN45_MOX}
\end{figure}

Both the chemical shift and branching ratio of the Am $N_{4,5}$ XAS spectra can be used to get information on the chemical state of Am in MOX. Fig.~\ref{AmN45_MOX} compares the Am $N_{4,5}$ XAS spectrum of the U$_{0.9}$Am$_{0.1}$O$_2$ sample with that of Am$_2$O$_3$. In terms of the chemical shift and relative $N_{4}$ intensity, both spectra are quite similar. Such a similarity suggests that americium in the the U$_{0.9}$Am$_{0.1}$O$_2$ sample is in the Am(III) state. That is in agreement with results of other studies of the U$_{1-x}$Am$_{x}$O$_2$ system \cite{Prieur,Vespa,Prieur02,Prieur03,Lebreton,Prieur04,Epifano,Mayer} and in particular of MOX with the same doped Am concentration (x=0.1). As to charge compensation of Am(III) in the UO$_2$ lattice, it was discussed that, instead of a creation of U(V), the electronic holes may be introduced in the O 2p band \cite{Kvashnina03} based on measurements and calculations of the O $K$ XAS spectrum of U$_{0.9}$Am$_{0.1}$O$_2$.

The observed difference in the relative $N_{4}$ intensity ratio of the Am $N_{4,5}$ XAS spectra between Am$_2$O$_3$ and AmO$_2$ is supported by the results of the AIM calculations of these spectra.
In the calculations, the ground state of AmO$_2$ was described as a linear combination of the $5f^{5}$,  $5f^{6}\underline{\upsilon}^{1}$ and $5f^{7}\underline{\upsilon}^{2}$ configurations, where $\underline{\upsilon}$ stands for an electronic hole in the O $2p$ band. The final state of the x-ray absorption process was represented by a combination of the $4d^{9}5f^{6}$, $4d^{9}5f^{7}\underline{\upsilon}^{1}$ and $4d^{9}5f^{8}\underline{\upsilon}^{2}$ configurations. In the limit of $V\rightarrow0$ (see the Computational details section), the difference between the configuration averaged energies for the ground state can be written as $E(5f^{n+1}\underline{\upsilon}^{1})-E(5f^{n})=\Delta$ and $E(5f^{n+2}\underline{\upsilon}^{2})-E(5f^{n+1}\underline{\upsilon}^{1}))=\Delta+U_{ff}$, where $\Delta$ is the Am $5f$-O $2p$ charge-transfer energy and $n$ is equal 5 in the AmO$_2$ case. $\Delta$ is taken as $\Delta=\epsilon_{5f}-\epsilon_{\upsilon}^0$. For the final state of the XAS process, the difference between the configuration averaged energies can be defined as $E(4d^{9}5f^{n+2}\underline{\upsilon}^{1})-E(4d^{9}5f^{n+1})=\Delta+U_{ff}-U_{fd}$ and $E(4d^{9}5f^{n+3}\underline{\upsilon}^{2})-E(4d^{9}5f^{n+2}\underline{\upsilon}^{1})=\Delta+2U_{ff}-U_{fd}$.

To reproduce the experimental Am $N_{4,5}$ XAS spectrum of AmO$_2$, the following values of the model parameters were used in the AIM calculations: $\Delta$=-0.25 eV, $U_{ff}$=6.2 eV, $U_{fd}$=7.1 eV, $V$=0.9 eV. These values are similar to those derived by Yamazaki and Kotani \textit{et al.} \cite{Yamazaki} from the AIM analysis of the Am $4f$ XPS spectrum of AmO$_2$. Since a combination of three configurations includes a very large number of the multiplet states, the value of $N$ parameter was set to one for simplicity in our AmO$_2$ calculations. The $F^k$ and $G^k$ integrals were scaled down to 80\% of their \textit{ab-initio} Hartree-Fock values calculated for the Am(IV) ion \cite{Butorin03}. There is a certain consensus to apply such a level of the Slater integral reduction for compounds. The values of Wybourne's crystal-field parameters ($B^{4}_{0}$=-0.84 eV and $B^{4}_{0}$=0.27 eV) for cubic symmetry were set to be the same as those in the Am $3d$-$4f$ RIXS calculations of AmO$_2$ using the crystal-field multiplet theory \cite{Butorin03}.

For Am$_2$O$_3$, the ground (final) state of the Am $N_{4,5}$ XAS process was described by a mixture of two configurations $5f^{6}$ and $5f^{7}\underline{\upsilon}^{1}$ ($4d^{9}5f^{7}$ and  $4d^{9}5f^{8}\underline{\upsilon}^{1}$) because the contribution of the $5f^{8}\underline{\upsilon}^{2}$ configuration is expected to be small due to significantly increased $\Delta$. In connection with that the $N$ parameter was set to 5 with $W$=2.0 eV. The other values of model parameters used in the AIM calculations for the Am $N_{4,5}$ XAS spectrum of Am$_2$O$_3$ were $\Delta$=6.5 eV, $U_{ff}$=5.7 eV, $U_{fd}$=6.0 eV and $V$=0.7 eV. The $F^k$ and $G^k$ integrals were also reduced to 80\% of their \textit{ab-initio} Hartree-Fock values calculated for the Am(III) ion \cite{Butorin04}. The crystal field was approximated by cubic symmetry with Wybourne's parameters set to $B^{4}_{0}$=-0.835 eV and $B^{4}_{0}$=0.100 eV. These parameter values were adopted from Ref.~\cite{Hubert} where they were derived using optical spectroscopy for Am(III) doped into the ThO$_2$ lattice.

It is interesting that the calculated Am $N_{4,5}$ XAS spectra (Fig.~\ref{AmN45_compare}) reproduce the observed small difference between Am$_2$O$_3$ and AmO$_2$ in the energy distance between the $N_{5}$ and $N_{4}$ lines which depends on the $4d$ spin-orbit splitting and the effect of the Am $5f$-O $2p$ hybridization. The difference in the $N_{4}$:$N_{5}$ intensity ratio between the Am $N_{4,5}$ XAS spectra of Am$_2$O$_3$ and AmO$_2$ seems to be somewhat larger in the calculations as compared with experiment (Fig.~\ref{AmN45_compare}), however, it can be in part connected to some difference in the core-hole broadening of the $N_{4}$ line between Am$_2$O$_3$ and AmO$_2$. For simplicity, the calculated Am $N_{4,5}$ XAS spectra were broadened with the $\Gamma_{m}$=2.0-eV Lorenzian \cite{Campbell} (besides the instrumental resolution approximated by the Gaussian), while it is expected that $\Gamma_{m}$ is somewhat larger for $N_{4}$ due to the $N_{4}\rightarrow{N_{5}}$ Coster-Kronig decay and interaction with the $N_{5}$ continuum. The smaller band gap in AmO$_2$ (Refs.~\cite{C_Suzuki,Noutack}) will result in a higher rate for the $N_{4}\rightarrow{N_{5}}$ Coster-Kronig decay, thus leading to a larger broadening of the $N_{4}$ line in AmO$_2$ as compared to that in Am$_2$O$_3$. However, it is not easy to obtain an exact estimate for that, since the transition probability for the valence electrons involved in the Coster-Kronig process varies throughout the valence band width.

To calculate the Am $O_{4,5}$ XAS spectrum of AmO$_2$, the same values of the AIM parameters were used. All the physical quantities and operators related to $4d$ were simply replaced in the Hamiltonian with those related to $5d$, so that $U_{fd}$ would stand for the $5d$ and $\epsilon_{d}$ would be the one-electron energy of the Am(IV) $5d$ level. The final state of the spectroscopic process was represented by a combination of the $5d^{9}5f^{6}$, $5d^{9}5f^{7}\underline{\upsilon}^{1}$ and $5d^{9}5f^{8}\underline{\upsilon}^{2}$ configurations. It has been shown \cite{Ogasawara,Kotani02} that XAS calculations at the An $5d$ edges require somewhat larger reduction of the \textit{ab-initio} Hartree-Fock atomic values of the $F^k$ and $G^k$ integrals, describing the $5d$-$5f$ interaction. Therefore, in our calculations the $F^{2,4,6}(5f,5f)$, $F^{2,4}(5d,5f)$, $G^{1,3,5}(5d,5f)$ integrals were scaled down to 80\%, 75\%, 65\%, respectively, of their \textit{ab-initio} values.

\begin{figure}
\centering
\includegraphics[width=0.8\columnwidth]{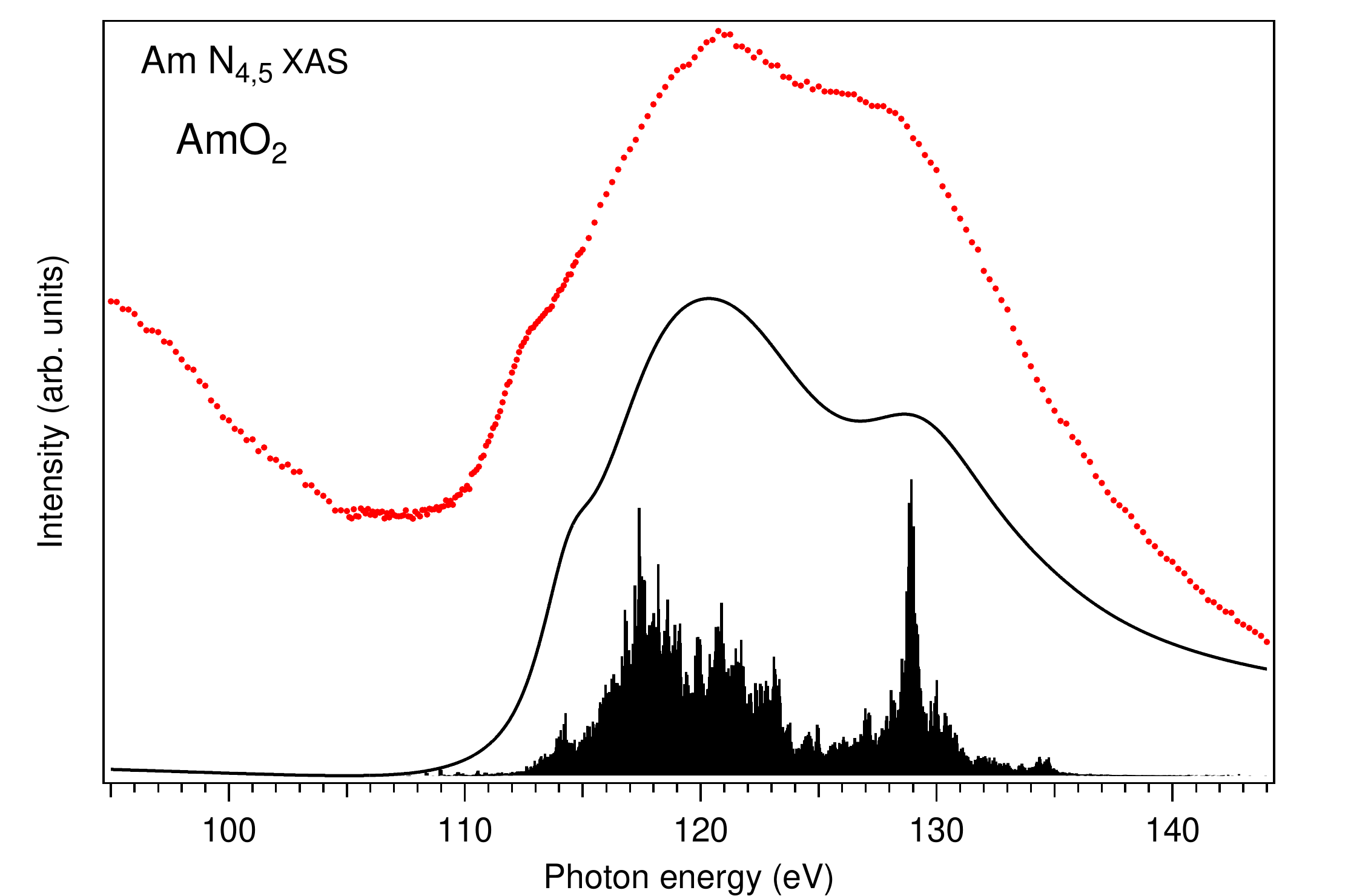}
\caption{Measured (red markers) and calculated (solid black line and poles) Am $O_{4,5}$ XAS spectra of AmO$_2$.}
\label{AmO45_XAS}
\end{figure}

The experimental Am $O_{4,5}$ XAS spectrum of AmO$_2$ displayed in Fig.~\ref{AmO45_XAS} appears to be significantly broadened by a short Am $5d$ core-hole lifetime as a result of super Coster-Kronig decay $\langle5d^{9}5f^{n+1}|1/r|5d^{10}5f^{n-1}\varepsilon{l}\rangle$ and other autoionization processes $\langle5d^{9}5f^{n+1}|1/r|5d^{10}5f^{n}5(s, p)^{-1}\varepsilon{l}\rangle$ and \\$\langle5d^{9}5f^{n+1}|1/r|5d^{9}5f^{n}\varepsilon{f}\rangle$. The $5d$ core-hole lifetime strongly varies throughout the $5d$ edge \cite{Butorin05} and substantially increases when going from the pre-threshold region to the main edge. For simplicity, the low-energy region of the calculated Am $O_{4,5}$ XAS spectrum up to 115.0 eV was broadened with the Lorentzian with $\Gamma_{m}$=1.0 eV and the rest of the spectrum was broadened with the Fano profile with $\Gamma_{m}$=3.0 eV (the instrumental resolution was also simulated by the corresponding Gaussian). The AIM calculations reproduce the experimental spectrum fairly well, thus supporting the choice of the model parameters and determined physical quantities based on these parameters. Note that there is some uncertainty on what function can used to fit a strongly diminishing-with-photon-energy background in the experimental spectrum in Fig~\ref{AmO45_XAS}, therefore the background was not subtracted and left as it is.

The AIM calculations of the An core-level XPS spectra offer a even stricter test of the choice of the AIM parameters because the spectra of the An oxides with strong hybridization between An and O states often reveal prominent charge-transfer satellites \cite{Yamazaki,Kotani02}. The energy positions of those satellites with respect to the main lines and their relative intensity allow for the more accurate determination of the values of the AIM parameters. Therefore, we performed the AIM calculations of Am $4f$ XPS spectra for both AmO$_2$ and Am$_2$O$_3$. The results of the calculations can be compared with available experimental data. The Am $4f$ XPS data were reported in a few publications for AmO$_2$ (Refs.~\cite{Teterin,Veal,Nevitt}) and Am$_2$O$_3$ (\cite{Mayer,Gouder,Finck,Nevitt}).

\begin{figure}
\centering
\includegraphics[width=0.8\columnwidth]{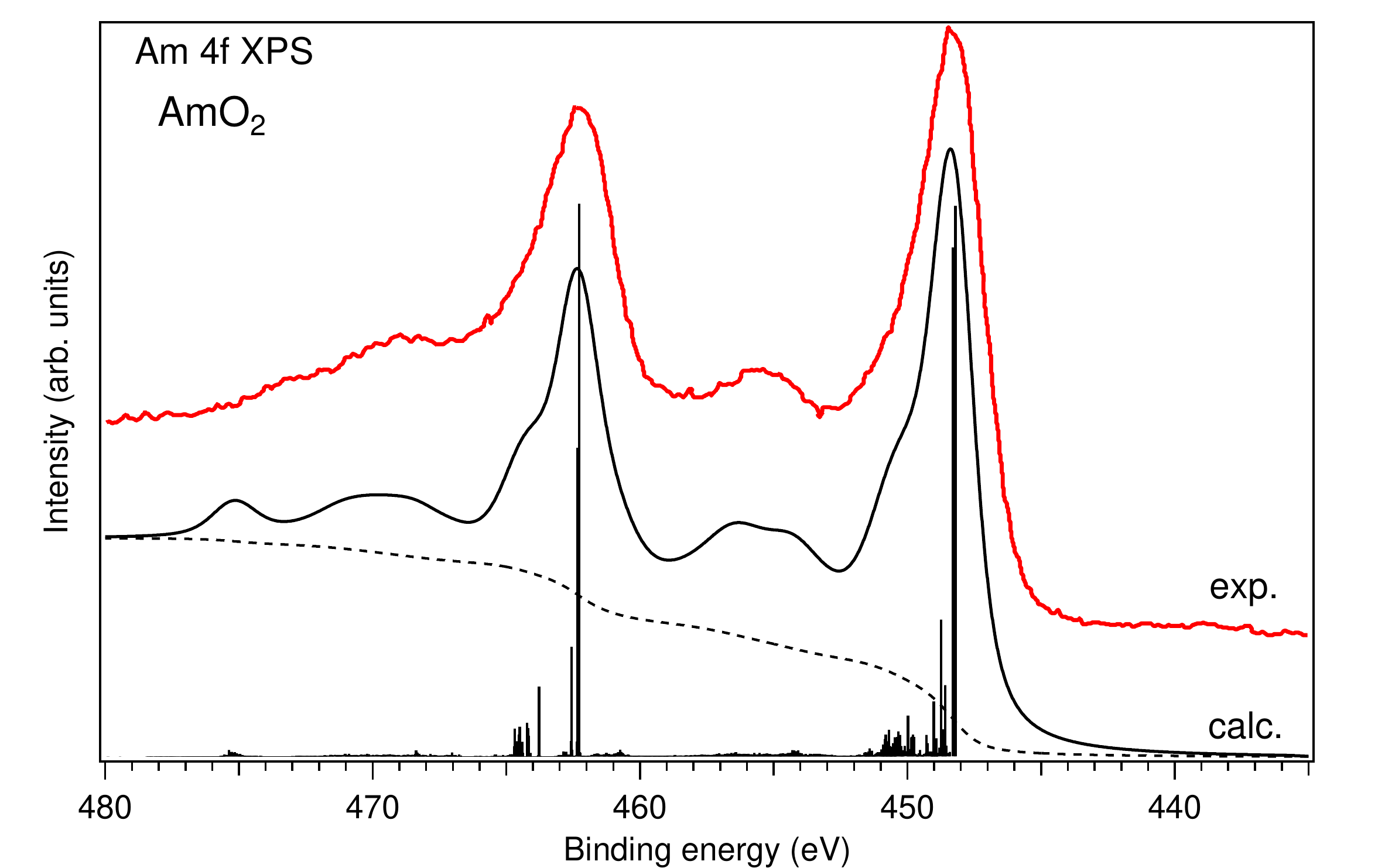}
\caption{Comparison of measured (adopted from Ref.~\cite{Nevitt}) and calculated Am $4f$ XPS spectra of AmO$_2$.}
\label{Am_4fXPS_AmO2}
\end{figure}

Fig.~\ref{Am_4fXPS_AmO2} compares the measured and calculated Am $4f$ XPS spectra of AmO$_2$. The experimental spectrum was adopted from Ref.~\cite{Nevitt}. The AIM calculations were performed for the same values of the model parameters and the same combination of the electronic configurations in the ground state as in case of the Am $N_{4,5}$ and $O_{4,5}$ XAS calculations for AmO$_2$. For the final state of the XPS process in AmO$_2$, a mixture of the $4f^{13}5f^{5}$, $4f^{13}5f^{6}\underline{\upsilon}^{1}$ and $4f^{13}5f^{7}\underline{\upsilon}^{2}$ configurations was used. The $U_{fc}$ value was set to 7.1 eV. In the limit of $V\rightarrow0$, the difference between the configuration averaged energies is described as $E(4f^{13}5f^{n+1}\underline{\upsilon}^{1})-E(4f^{13}5f^{n})=\Delta-U_{fc}$ and $E(4f^{13}5f^{n+2}\underline{\upsilon}^{2})-E(4f^{13}5f^{n+1}\underline{\upsilon}^{1})=\Delta+U_{ff}-U_{fc}$. The only difference here, within the same computational approach, from the cases of Am $N_{4,5}$ and $O_{4,5}$ XAS is the scaling amount of the \textit{ab-initio} $G^k$ integrals, which will be discussed later.

The experimental Am $4f$ XPS of AmO$_2$ in Fig.~\ref{Am_4fXPS_AmO2} contains the main $4f_{7/2}$ and $4f_{5/2}$ lines at around 448.2 eV and 462.3 eV, respectively, and is indeed characterized by the presence of the pronounced Am $5f$-O $2p$ charge-transfer satellites located at $\sim$455.3 eV and $\sim$469.4 eV, respectively. In addition, the hint of another structure at $\sim$473.2 eV can be recognized in Fig.~\ref{Am_4fXPS_AmO2} while this structure is more clearly resolved in Ref.~\cite{Teterin}. The AIM calculations of the Am $4f$ XPS spectrum of AmO$_2$ reproduce the experimental structures quite well, except for the $\sim$473.2-eV structure. The latter in the calculated spectrum is located at the biding energies around 475 eV and is associated with the contribution of the $4f^{13}5f^{7}\underline{\upsilon}^{2}$ configuration. Nevertheless, the energy position and the relative intensity of this structure is anticipated to be in better agreement with experiment when more electronic configurations ($4f^{13}5f^{7+n}\underline{\upsilon}^{2+n}$) are included in the calculations. Due to a huge number of the involved multiplets and a high demand on the computational resources, the present calculations were limited to the current number of the electronic configurations.

\begin{figure}
\centering
\includegraphics[width=0.8\columnwidth]{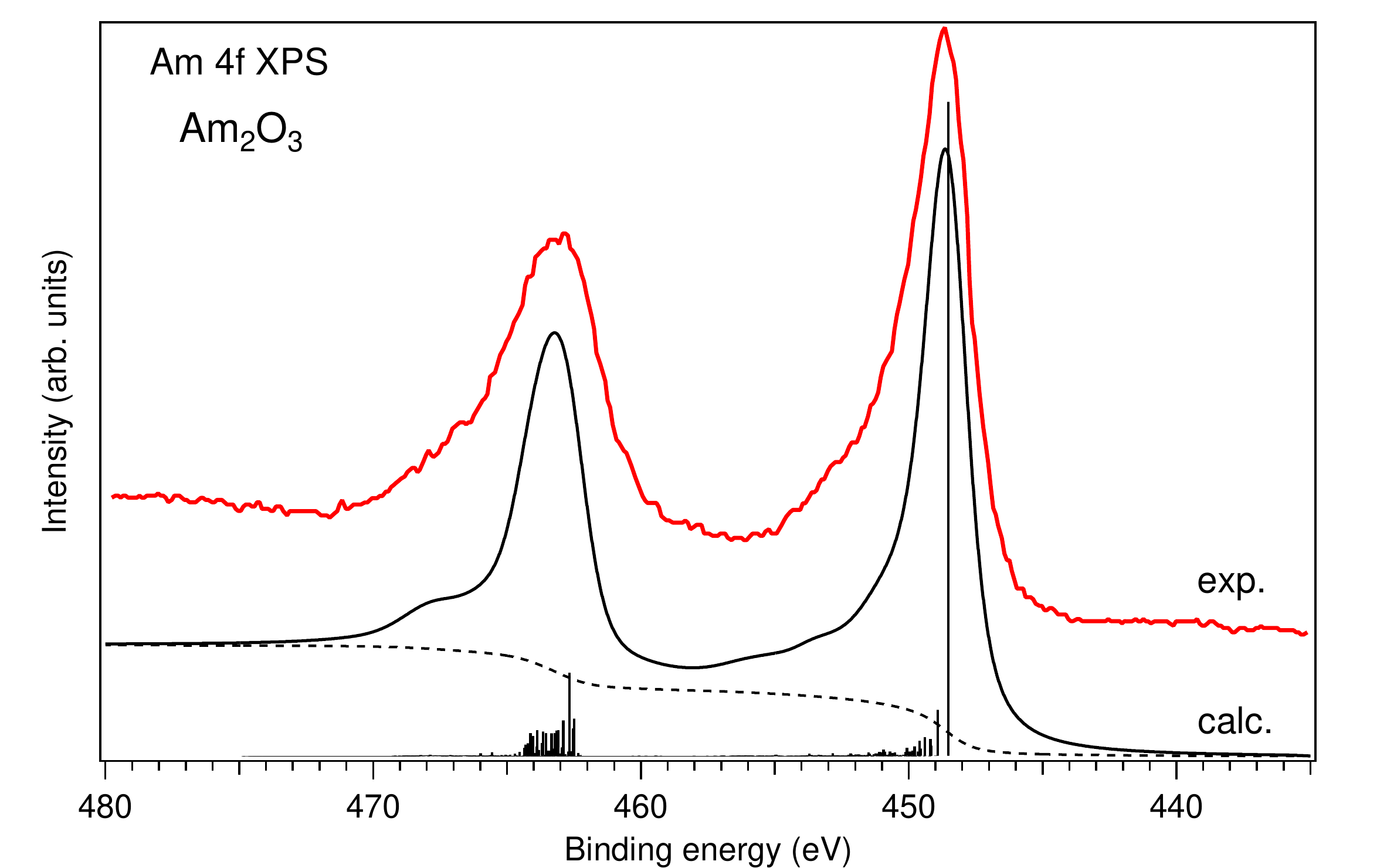}
\caption{Comparison of measured (adopted from Ref.~\cite{Nevitt}) and calculated Am $4f$ XPS spectra of Am$_2$O$_3$.}
\label{Am_4fXPS_Am2O3}
\end{figure}

Fig.~\ref{Am_4fXPS_Am2O3} compares the calculated Am $4f$ XPS spectrum of Am$_2$O$_3$ to the experimental data adopted from Ref.~\cite{Nevitt}. Again, the AIM calculations were performed for the same values of model parameters as in case of the Am $N_{4,5}$ XAS calculations for Am$_2$O$_3$ except for the scaling of the \textit{ab-initio} $G^k$ integrals. The final state was described by a mixture of the $4f^{13}5f^{6}$ and $4f^{13}5f^{7}\underline{\upsilon}^{1}$ with the $U_{fc}$ value equal to 6.0 eV. As a result, the observed agreement between experiment and calculations is quite good, thereby indicating the correct choice of the values for the AIM model parameters. Note, that the energy scales for the experimental Am $4f$ XPS of AmO$_2$ and Am$_2$O$_3$ were kept exactly the same as shown in  Ref.~\cite{Nevitt} while the corresponding calculated spectra were aligned with the experimental ones.

\begin{figure}
\centering
\includegraphics[width=0.8\columnwidth]{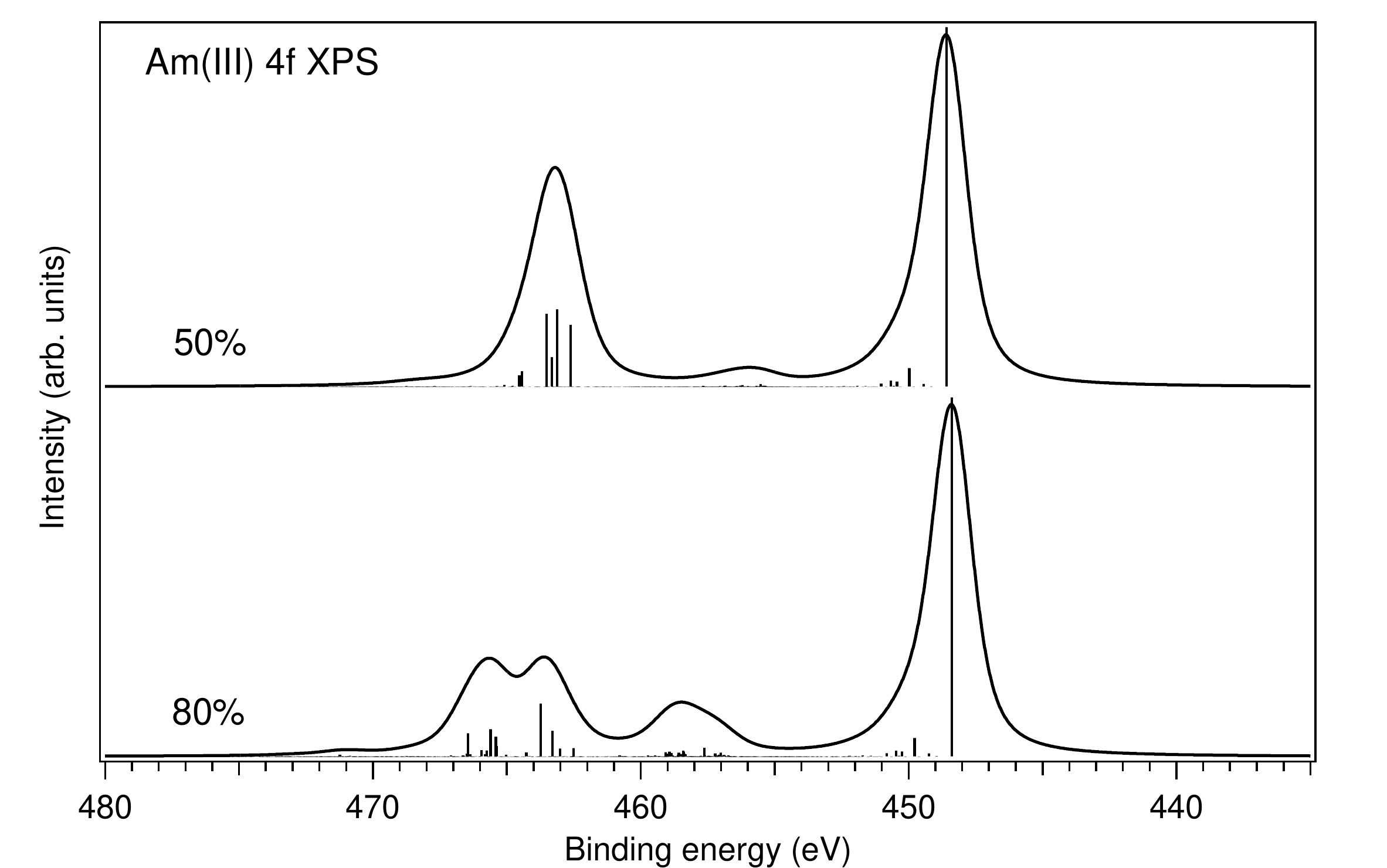}
\caption{Calculated $4f$ XPS spectra of the Am(III) ion with the reduction of $G^k$ integrals to 80\% and 50\% of their \textit{ab-initio} Hartree-Fock values, respectively.}
\label{AmIII_4fXPS}
\end{figure}

\begin{figure}
\centering
\includegraphics[width=0.8\columnwidth]{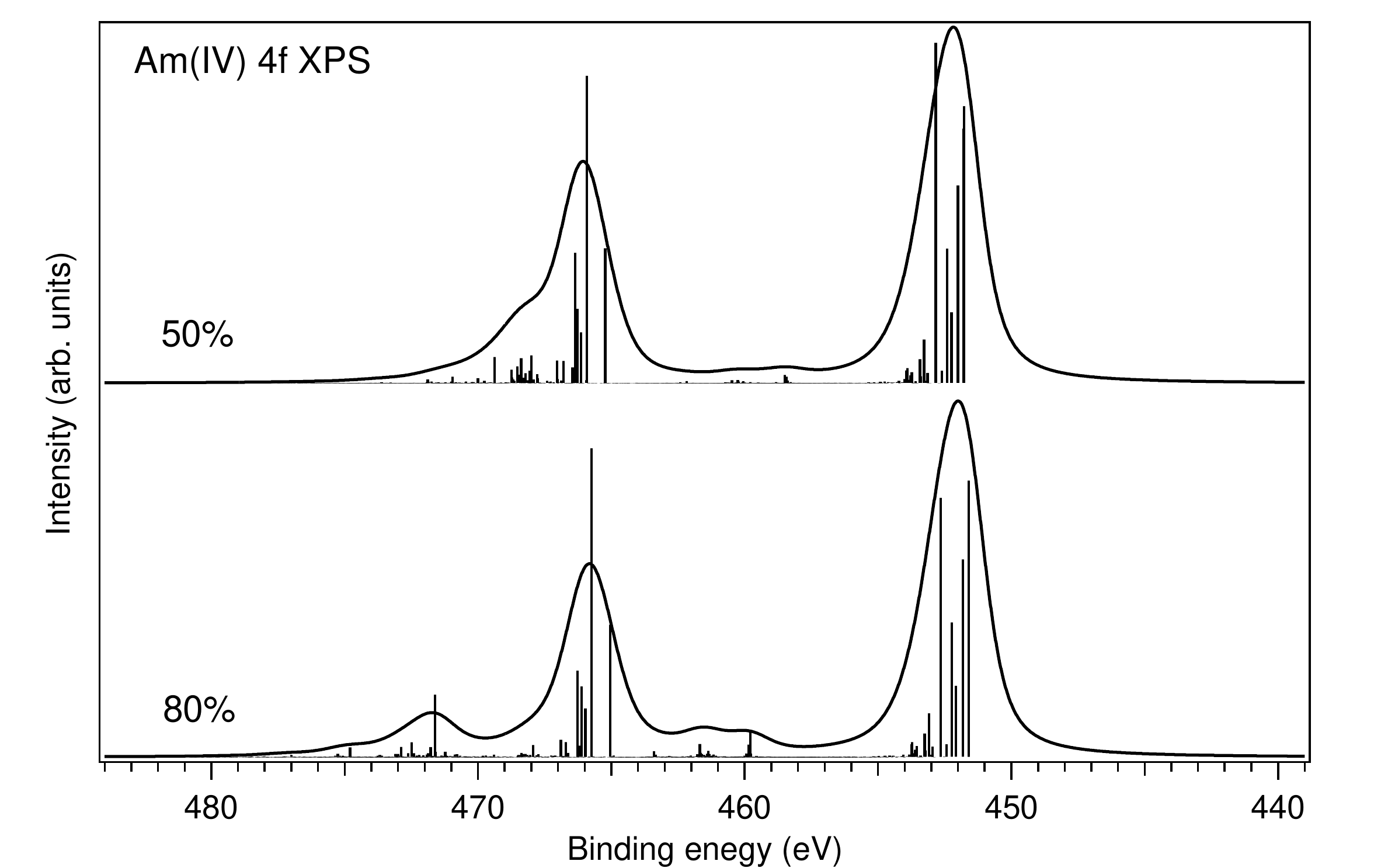}
\caption{Calculated $4f$ XPS spectra of the Am(IV) ion with the reduction of $G^k$ integrals to 80\% and 50\% of their \textit{ab-initio} Hartree-Fock values, respectively.}
\label{AmIV_4fXPS}
\end{figure}

In solids, the value of the $G^0$ integral is expected to be significantly screened as compared to that for a free ion. In the case of the Am $4f$ XPS spectrum, it is clear that the conventional reduction of $G^k$ integrals to 80\% of their \textit{ab-initio} Hartree-Fock values does not fully account for such screening. Figs.~\ref{AmIII_4fXPS} and \ref{AmIV_4fXPS} display the results of the atomic multiplet calculations of the $4f$ XPS spectra for the Am(III) and Am(IV) ions, respectively, with different scaling of the $G^k$ integrals while the reduction of the $F^k$ integrals were kept to 80\% of their \textit{ab-initio} Hartree-Fock values. As one can see in Fig.~\ref{AmIII_4fXPS}, the Am(III) $4f$ XPS spectrum with the $G^k$ reduction to 80\% reveals an intense structure at $\sim$459.3 eV and a double-peak $4f_{5/2}$ line (at $\sim$464.4 eV and $\sim$466.5 eV) which are not observed in the experimental Am $4f$ XPS spectrum of Am$_2$O$_3$ at all. A significant $G^k$ scaling down to 50\% of their \textit{ab-initio} Hartree-Fock values is required for these $4f$ XPS extra-structures to be significantly reduced (see Fig.~\ref{AmIII_4fXPS}). A similar situation was found for Am(IV) (Fig.~\ref{AmIV_4fXPS}). As a result of this exercise, the Am $4f$ XPS spectra of AmO$_2$ and Am$_2$O$_3$ in Figs.~\ref{Am_4fXPS_AmO2} and \ref{Am_4fXPS_Am2O3} were calculated with the 50\% reduction of the $G^k$ integrals. Note, that in this case, the main effect comes from the scaling of $G^0$, the reduction of $G^{2,4,6}$ integrals does not affect the shape of the Am $4f$ XPS spectrum much.

\section*{Conclusions}
AIM calculations take into account all the important interactions to characterize chemical bonding. For AmO$_2$, the ground state in cubic symmetry is $\Gamma_7$ and it does not change when the Am $5f$-O $2p$ hybridization is taken into account in the AIM calculations for AmO$_2$. Note, that a possible multipolar magnetic order \cite{Suzuki,Noutack} was disregarded in our calculations. The contributions of the $5f^{5}$, $5f^{6}\underline{\upsilon}^{1}$ and $5f^{7}\underline{\upsilon}^{2}$ configurations in the ground state of AmO$_2$ were calculated to be 34\%, 59\% and 7\%, respectively. This results in $5f$ occupancy $n_f$=5.73 electrons and indicates a significant covalency of the Am $5f$-O $2p$ bonds. For Am$_2$O$_3$, the contributions of the $5f^{6}$ and $5f^{7}\underline{\upsilon}^{1}$ configurations in the $\Gamma_1$ ground state were found to be 95\% and 5\%, respectively, thus leading to $n_f$=6.05 electrons. Regarding the $\Delta/U_{ff}$ ratio \cite{Zaanen}, AmO$_2$ and can be classified as the charge-transfer compound while Am$_2$O$_3$ is rather a Mott-Hubbard system.

\section*{Methods}
\subsection*{Experimental}
One of the Am oxide samples used for measurements was fabricated by technique used to prepare radionuclide counting plates at the Lawrence Berkeley National Laboratory (LBNL; see the "Preparation of counting sources" subsection in Ref. \cite{Moody}). The counting plate was prepared from an aqueous solution of about 2.0 mM Am-243 (better than 99.6\% Am-243 by mass) in 0.1 M HCl that was delivered by micropipette techniques to an area of ~4 mm$^2$ area on a high purity Pt substrate (25.4 mm diameter). The aqueous droplets were allowed dry leaving a residue that was ring-shaped. This was followed by inductive heating to nearly 700 $^o$C under atmosphere to oxidize the material and fixing the material to the Pt substrate to preclude loss when placed in the UHV spectrometer chamber during the measurement. This process is expected to yield the Am oxide sesquioxide with an approximate composition of Am$_2$O$_3$ (Ref.\cite{Kvashnina_Cm}). The counting plated was trimmed to 3 mm x 3 mm around the center and mounted on the sample with conductive tape as described below. The Am sample taken to the Advanced Light Source (ALS) was close to 1 $\mu$g of Am-243, approaching the safety limit of 200 nCi for Am-243.

The measured relative intensity ratio of the O $K\alpha$ ($2p\rightarrow1s$ transitions) and Am $N_{5}$-$N_{6,7}$ ($4f\rightarrow4d$ transitions) or Am $O_{4,5}$-$P_{2,3}$ ($6p\rightarrow5d$ transitions) lines (from low-energy-resolution overview spectra) corresponded to the Am$_2$O$_3$ oxide as compared to AmO$_2$. Furthermore, the Am $5d$-$5f$ resonant inelastic x-ray scattering (RIXS) spectra at the Am $5d$ edge of this sample revealed the presence of a strong RIXS transition at the energy loss of $\sim$300 meV, which is characteristic for Am(III). A specially designed sample holder, which is described in Refs. \cite{Butorin05,Shuh}, was used for the Am$_2$O$_3$ sample during the measurements. It is essentially a cylindrical can with slots for incoming and outgoing radiation. The
sample is attached to the slab inside the can just behind the slot. Due to such a design, the sample holder served as a catch tray for material that might come loose during handling and the measurements, thus ensuring that no contamination will be left in the experimental chamber after the experiment.

The U$_{0.9}$Am$_{0.1}$O$_2$ sample was prepared by a conventional powder metallurgical method. Required amounts of the depleted-uranium dioxide and (Am-241)O$_2$ powders were weighed for the preparation of U$_{0.9}$Am$_{0.1}$O$_2$ and mixed in a ball mill using tungsten balls. The resulting powder was pressed into a pellet at 40 MPa after adding the organic binder. To remove the organic binder, the pellet was heated at 800 C for 2.5 h in a reducing atmosphere. Sintering of the pellet was performed at 1700 $^o$C for 3 h under an Ar atmosphere containing 5\% H$_2$. Both heating and cooling rates were 200 $^o$C/h. The oxygen potentials of the sintering atmosphere were adjusted by adding moisture. The prepared sample was characterized by x-ray diffraction. For x-ray spectroscopic measurements in the soft x-ray range at the synchrotron radiation laboratory, a tiny fraction of the prepared pellet was used to mount the sample in the closed source experimental system, described in Ref. \cite{Modin}. This closed source experimental system was also used for a flake of AmO$_2$ with the size of $\sim$0.3x0.3 mm$^2$. Instead of a Si$_3$N$_4$ window, a diamond window with the thickness of 100 nm was installed to provide the higher x-ray transmission in the energy range of the Am $5d$ edge. A slight improvement to the experimental system was made to measure the drain current on the sample.

The measurements in the energy range of the Am $N_{4,5}$ ($4d\rightarrow5f,7p$ transitions) and $O_{4,5}$ ($5d\rightarrow5f,7p$ transitions) edges of the AmO$_2$ and U$_{0.9}$Am$_{0.1}$O$_2$ samples were performed at beamline 5.3.1 of the MAXlab \cite{Denecke}. The Am $4d$ and $5d$ XAS data were measured in the total electron yield (TEY) mode using the drain current on the sample. The incidence angle of the incoming photons was close to 90$^\circ$ to the surface of the sample. The monochromator resolution was set to $\sim$600 meV at 840 eV during measurements at the Am $4d$ edges and to $\sim$50 meV at 115 eV during measurements at the Am $5d$ edges. The Am $N_{4,5}$ XAS spectra of the Am$_2$O$_3$ sample were recorded at beamline 7.0.1 of ALS \cite{Warwick} of LBNL with the same energy resolution and at the same incidence angle as in measurements on AmO$_2$ and U$_{0.9}$Am$_{0.1}$O$_2$. To optimize the placement of the photon beam on small samples, a camera with the zoom option was taken advantage of in both experiments at MAXlab and ALS. The camera was attached to the flange window on the analyzing chamber. To bring the Am $N_{4,5}$ XAS spectra measured at MAXlab and ALS to the same energy scale, the Ni $L_{2,3}$ XAS spectra of a Ni foil were recorded in both cases.

\subsection*{Computational Details}
AIM \cite{Anderson} was used for the calculations which included the $5f$ and core $4d$($5d$) or $4f$ states on a single actinide ion and the O $2p$ states. The calculations were performed in a manner described in Refs. \cite{Kotani,Butorin06,Nakazawa}.

The total Hamiltonian of a system can be written as
\begin{eqnarray}
H&=&\epsilon_{5f}\sum_{\gamma} a^{\dag}_{5f}(\gamma)a_{5f}(\gamma) \nonumber  \\
       &+&
\epsilon_{d}\sum_{\mu} a^{\dag}_{d}(\mu)a_{d}(\mu) \nonumber  \\
       &+&
\epsilon_{4f}\sum_{\lambda} a^{\dag}_{4f}(\lambda)a_{4f}(\lambda) \nonumber  \\
       &+&
\sum_{\sigma,\gamma} \epsilon_{\upsilon}(\sigma)a^{\dag}_{\upsilon}(\sigma,\gamma)a_{\upsilon}(\sigma,\gamma) \nonumber  \\
       &+& U_{ff}\sum_{\gamma>\gamma^{\prime}}a^{\dag}_{5f}(\gamma)a_{5f}(\gamma)a^{\dag}_{5f}(\gamma^{\prime})a_{5f}(\gamma^{\prime})  \nonumber\\
       &-&
       U_{fd}\sum_{\gamma,\mu}a^{\dag}_{5f}(\gamma)a_{5f}(\gamma)a^{\dag}_{3d}(\mu)a_{3d}(\mu) \nonumber\\
       &-&
       U_{fc}\sum_{\gamma,\lambda}a^{\dag}_{5f}(\gamma)a_{5f}(\gamma)a^{\dag}_{4f}(\lambda)a_{4f}(\lambda) \nonumber\\
       &+&
       \frac{V}{\sqrt{N}}\sum_{\sigma,\gamma} [(a^{\dag}_{\upsilon}(\sigma,\gamma)a_{5f}(\gamma) + a^{\dag}_{5f}(\gamma)a_{\upsilon}(\sigma,\gamma)] \nonumber\\
       &+&
       H_{multiplet},
\end{eqnarray}
where $\epsilon_{5f}$, $\epsilon_{d}$, $\epsilon_{4f}$ and $\epsilon_{\upsilon}$ are one-electron energies of actinide $5f$, core $4d$($5d$) and $4f$ levels and valence band, respectively, and $a^{\dag}_{5f}(\gamma)$, $a^{\dag}_{d}(\mu)$, $a^{\dag}_{4f}(\lambda)$, $a^{\dag}_{\upsilon}(\sigma,\gamma)$ are electron creation operators at these levels with combined indexes $\gamma$, $\mu$ and $\lambda$ to represent the spin and orbital states of the $5f$, $4d$($5d$) and $4f$ and valence-band electrons, $\sigma$ is the index of the $N$ discrete energy levels in the O $2p$ band (bath states). $U_{fd}$ and $U_{fc}$ are the $4d$($5d$) and $4f$ core hole potentials, respectively, acting on the $5f$ electron. $V$ is the hybridization term between actinide $5f$ states and states of the O $2p$ band. The $\epsilon_{\upsilon}(\sigma)$ is represented by the $N$ discrete levels/bath states in the form
\begin{eqnarray}
\epsilon_{\upsilon}(\sigma)=\epsilon_{\upsilon}^0-\frac{W}{2}+\frac{W}{N}(\sigma-\frac{1}{2}),~\sigma=1,...,N,
\end{eqnarray}
where $\epsilon_{\upsilon}^0$ and $W$ are the center and width of the O $2p$ band, respectively. $H_{multiplet}$ represents the electrostatic ($F^k$), exchange ($G^k$) and spin-orbit interactions for the actinide ion and the applied crystal field \cite{Butorin07,Butorin03,Butorin04}.

The isotropic XAS spectra at the Am $N_{4,5}$ edges were calculated using the equation
\begin{eqnarray}
I_{XAS}(\omega) = \sum_{m} | \langle m | D | g \rangle |^{2} \frac{\Gamma_{m}/\pi}{(E_{m}-E_{g}-\omega)^{2}+\Gamma_{m}^{2}},
\end{eqnarray}
where $| g \rangle$ and $| m \rangle$ are the ground and XAS final states of the spectroscopic process with energies $E_{g}$ and $E_{m}$, respectively. $D$ is the operator for the optical dipole transition with the incident photon energy represented by $\omega$ and lifetime broadening  $\Gamma_{m}$ of the final state in terms of half-width at half-maximum (HWHM).

The Am $4f$ XPS spectra were calculated using the following equation
\begin{eqnarray}
I_{XPS}(E_{B}) = \sum_{f} | \langle f | a_c | g \rangle |^{2} \frac{\Gamma_{f}/\pi}{(E_{f}-E_{g}-E_{B})^{2}+\Gamma_{f}^{2}},
\end{eqnarray}
where $| g \rangle$ and $| f \rangle$ are the ground and XPS final states of the spectroscopic process with energies $E_{g}$ and $E_{f}$, respectively. $E_{B}$ is the binding energy, and $a_c$ is the annihilation operator of a core electron and $\Gamma_{f}$ is a lifetime broadening of the XPS final state in terms of HWHM.

The \textit{ab-initio} values of Slater integrals $F^{2,4,6}(5f,5f)$, $F^{2,4}(4d,5f)$, $F^{2,4,6}(4f,5f)$, $G^{1,3,5}(4d,5f)$, $G^{0,2,4,6}(4f,5f)$, spin-orbit coupling constants $\zeta(5f)$, $\zeta(4d)$, $\zeta(4f)$ and matrix elements were obtained with the TT-MULTIPLETS package which combines Cowan's atomic multiplet program \cite{Cowan} (based on the Hartree-Fock method with relativistic corrections) and Butler's point-group program \cite{Butler}, which were modified by Thole \cite{Thole}, as well as the charge-transfer program written by Thole and Ogasawara.
To compare with the experimental data, it is usually necessary to uniformly shift the calculated spectra on the photon energy scale because it is difficult to accurately reproduce the absolute energies in this type of calculations.

\section*{Acknowledgement}
S.M.B acknowledges the support from the Swedish Research Council (research grant 2017-06465). The computations and data handling were enabled by resources provided by the Swedish National Infrastructure for Computing (SNIC) at National Supercomputer Centre at Link\"{o}ping University partially funded by the Swedish Research Council through grant agreement no. 2018-05973.

This work was supported in part by the Director, Office of Science, Office of Basic Energy Sciences, Division of Chemical Sciences, Geosciences, and Biosciences Heavy Elements Chemistry program (DKS) of the U.S. Department of Energy at Lawrence Berkeley National Laboratory under Contract No. DE-AC02-05CH11231. This research used resources of the Advanced Light Source, a U.S. DOE Office of Science User Facility under contract No. DE-AC02-05CH11231. The Am-243 used in this work was supplied by the U.S. DOE through the transplutonium element production facilities at ORNL.

\section*{Author contributions statement}
S.M.B. conceived, planned and conducted the experiments, performed the calculations and analyzed the results, D.K.S. planned and conducted the experiments, prepared and characterized the samples. The authors reviewed the manuscript.

\section*{Additional information}
\subsubsection*{Competing interests}
The authors declare no competing interests.

\subsubsection*{Data availability}
The datasets generated during and/or analysed during the current study are available from the corresponding author on reasonable request.



\bibliography{Am_oxides}

\begin{thebibliography}{10}
\urlstyle{rm}
\expandafter\ifx\csname url\endcsname\relax
  \def\url#1{\texttt{#1}}\fi
\expandafter\ifx\csname urlprefix\endcsname\relax\def\urlprefix{URL }\fi
\expandafter\ifx\csname doiprefix\endcsname\relax\def\doiprefix{DOI: }\fi
\providecommand{\bibinfo}[2]{#2}
\providecommand{\eprint}[2][]{\url{#2}}

\bibitem{Wiss}
\bibinfo{author}{Wiss, T.} \emph{et~al.}
\newblock \bibinfo{journal}{\bibinfo{title}{Investigation on the use of
  {Americium} {Oxide} for {Space} {Power} {Sources}: {Radiation} {Damage}
  {Studies}}}.
\newblock {\emph{\JournalTitle{E3S Web Conf.}}} \textbf{\bibinfo{volume}{16}},
  \bibinfo{pages}{05004}, \doiprefix\url{10.1051/e3sconf/20171605004}
  (\bibinfo{year}{2017}).

\bibitem{Vigier}
\bibinfo{author}{Vigier, J.-F.} \emph{et~al.}
\newblock \bibinfo{journal}{\bibinfo{title}{Optimization of {Uranium}-{Doped}
  {Americium} {Oxide} {Synthesis} for {Space} {Application}}}.
\newblock {\emph{\JournalTitle{Inorg. Chem.}}} \textbf{\bibinfo{volume}{57}},
  \bibinfo{pages}{4317--4327}, \doiprefix\url{10.1021/acs.inorgchem.7b03148}
  (\bibinfo{year}{2018}).

\bibitem{Wiss02}
\bibinfo{author}{Wiss, T.} \emph{et~al.}
\newblock \bibinfo{journal}{\bibinfo{title}{{TEM} study of alpha-damaged
  plutonium and americium dioxides}}.
\newblock {\emph{\JournalTitle{J. Mater. Res.}}} \textbf{\bibinfo{volume}{30}},
  \bibinfo{pages}{1544--1554}, \doiprefix\url{10.1557/jmr.2015.37}
  (\bibinfo{year}{2015}).

\bibitem{Nishi}
\bibinfo{author}{Nishi, T.} \emph{et~al.}
\newblock \bibinfo{journal}{\bibinfo{title}{Local and electronic structure of
  {Am$_2$O$_3$} and {AmO$_2$} with {XAFS} spectroscopy}}.
\newblock {\emph{\JournalTitle{Journal of Nuclear Materials}}}
  \textbf{\bibinfo{volume}{401}}, \bibinfo{pages}{138--142},
  \doiprefix\url{10.1016/j.jnucmat.2010.04.011} (\bibinfo{year}{2010}).

\bibitem{Prieur}
\bibinfo{author}{Prieur, D.} \emph{et~al.}
\newblock \bibinfo{journal}{\bibinfo{title}{Local {Structure} and {Charge}
  {Distribution} in {Mixed} {Uranium}–{Americium} {Oxides}: {Effects} of
  {Oxygen} {Potential} and {Am} {Content}}}.
\newblock {\emph{\JournalTitle{Inorg. Chem.}}} \textbf{\bibinfo{volume}{50}},
  \bibinfo{pages}{12437--12445}, \doiprefix\url{10.1021/ic200910f}
  (\bibinfo{year}{2011}).

\bibitem{Vespa}
\bibinfo{author}{Vespa, M.}, \bibinfo{author}{Rini, M.},
  \bibinfo{author}{Spino, J.}, \bibinfo{author}{Vitova, T.} \&
  \bibinfo{author}{Somers, J.}
\newblock \bibinfo{journal}{\bibinfo{title}{Fabrication and characterization of
  {(U,Am)O$_{2-x}$} transmutation targets}}.
\newblock {\emph{\JournalTitle{Journal of Nuclear Materials}}}
  \textbf{\bibinfo{volume}{421}}, \bibinfo{pages}{80--88},
  \doiprefix\url{10.1016/j.jnucmat.2011.11.055} (\bibinfo{year}{2012}).

\bibitem{Prieur02}
\bibinfo{author}{Prieur, D.} \emph{et~al.}
\newblock \bibinfo{journal}{\bibinfo{title}{Reactive sintering of
  {U$_{1-y}$Am$_y$O$_{2{\pm}x}$} in overstoichiometric conditions}}.
\newblock {\emph{\JournalTitle{Journal of the European Ceramic Society}}}
  \textbf{\bibinfo{volume}{32}}, \bibinfo{pages}{1585--1591},
  \doiprefix\url{10.1016/j.jeurceramsoc.2011.12.017} (\bibinfo{year}{2012}).

\bibitem{Prieur03}
\bibinfo{author}{Prieur, D.} \emph{et~al.}
\newblock \bibinfo{journal}{\bibinfo{title}{Comparative {XRPD} and {XAS} study
  of the impact of the synthesis process on the electronic and structural
  environments of uranium–americium mixed oxides}}.
\newblock {\emph{\JournalTitle{Journal of Solid State Chemistry}}}
  \textbf{\bibinfo{volume}{230}}, \bibinfo{pages}{8--13},
  \doiprefix\url{10.1016/j.jssc.2015.03.037} (\bibinfo{year}{2015}).

\bibitem{Lebreton}
\bibinfo{author}{Lebreton, F.} \emph{et~al.}
\newblock \bibinfo{journal}{\bibinfo{title}{Peculiar {Behavior} of
  {(U,Am)O$_{2-x}$} {Compounds} for {High} {Americium} {Contents} {Evidenced}
  by {XRD}, {XAS}, and {Raman} {Spectroscopy}}}.
\newblock {\emph{\JournalTitle{Inorg. Chem.}}} \textbf{\bibinfo{volume}{54}},
  \bibinfo{pages}{9749--9760}, \doiprefix\url{10.1021/acs.inorgchem.5b01357}
  (\bibinfo{year}{2015}).

\bibitem{Prieur04}
\bibinfo{author}{Prieur, D.} \emph{et~al.}
\newblock \bibinfo{journal}{\bibinfo{title}{Melting behaviour of
  americium-doped uranium dioxide}}.
\newblock {\emph{\JournalTitle{The Journal of Chemical Thermodynamics}}}
  \textbf{\bibinfo{volume}{97}}, \bibinfo{pages}{244--252},
  \doiprefix\url{10.1016/j.jct.2016.02.003} (\bibinfo{year}{2016}).

\bibitem{Epifano}
\bibinfo{author}{Epifano, E.} \emph{et~al.}
\newblock \bibinfo{journal}{\bibinfo{title}{Extreme multi-valence states in
  mixed actinide oxides}}.
\newblock {\emph{\JournalTitle{Commun Chem}}} \textbf{\bibinfo{volume}{2}},
  \bibinfo{pages}{59}, \doiprefix\url{10.1038/s42004-019-0161-0}
  (\bibinfo{year}{2019}).

\bibitem{Butorin}
\bibinfo{author}{Butorin, S.~M.} \emph{et~al.}
\newblock \bibinfo{journal}{\bibinfo{title}{Local {Symmetry} {Effects} in
  {Actinide} 4f {X}-ray {Absorption} in {Oxides}}}.
\newblock {\emph{\JournalTitle{Anal. Chem.}}} \textbf{\bibinfo{volume}{88}},
  \bibinfo{pages}{4169--4173}, \doiprefix\url{10.1021/acs.analchem.5b04380}
  (\bibinfo{year}{2016}).

\bibitem{Kvashnina}
\bibinfo{author}{Kvashnina, K.~O.}, \bibinfo{author}{Butorin, S.~M.},
  \bibinfo{author}{Martin, P.} \& \bibinfo{author}{Glatzel, P.}
\newblock \bibinfo{journal}{\bibinfo{title}{Chemical {State} of {Complex}
  {Uranium} {Oxides}}}.
\newblock {\emph{\JournalTitle{Phys. Rev. Lett.}}}
  \textbf{\bibinfo{volume}{111}}, \bibinfo{pages}{253002},
  \doiprefix\url{10.1103/PhysRevLett.111.253002} (\bibinfo{year}{2013}).

\bibitem{Kvashnina02}
\bibinfo{author}{Kvashnina, K.}, \bibinfo{author}{Kvashnin, Y.} \&
  \bibinfo{author}{Butorin, S.}
\newblock \bibinfo{journal}{\bibinfo{title}{Role of resonant inelastic {X}-ray
  scattering in high-resolution core-level spectroscopy of actinide
  materials}}.
\newblock {\emph{\JournalTitle{Journal of Electron Spectroscopy and Related
  Phenomena}}} \textbf{\bibinfo{volume}{194}}, \bibinfo{pages}{27--36},
  \doiprefix\url{10.1016/j.elspec.2014.01.016} (\bibinfo{year}{2014}).

\bibitem{Butorin02}
\bibinfo{author}{Butorin, S.~M.}, \bibinfo{author}{Kvashnina, K.~O.},
  \bibinfo{author}{Vegelius, J.~R.}, \bibinfo{author}{Meyer, D.} \&
  \bibinfo{author}{Shuh, D.~K.}
\newblock \bibinfo{journal}{\bibinfo{title}{High-resolution {X}-ray absorption
  spectroscopy as a probe of crystal-field and covalency effects in actinide
  compounds}}.
\newblock {\emph{\JournalTitle{Proc. Natl. Acad. Sci. U.S.A.}}}
  \textbf{\bibinfo{volume}{113}}, \bibinfo{pages}{8093--8097},
  \doiprefix\url{10.1073/pnas.1601741113} (\bibinfo{year}{2016}).

\bibitem{Butorin03}
\bibinfo{author}{Butorin, S.~M.}
\newblock \bibinfo{journal}{\bibinfo{title}{3d-4f {Resonant} {Inelastic}
  {X}-ray {Scattering} of {Actinide} {Dioxides}: {Crystal}-{Field} {Multiplet}
  {Description}}}.
\newblock {\emph{\JournalTitle{Inorg. Chem.}}} \textbf{\bibinfo{volume}{59}},
  \bibinfo{pages}{16251--16264}, \doiprefix\url{10.1021/acs.inorgchem.0c02032}
  (\bibinfo{year}{2020}).

\bibitem{Butorin04}
\bibinfo{author}{Butorin, S.~M.}
\newblock \bibinfo{journal}{\bibinfo{title}{Advanced x-ray spectroscopy of
  actinide trichlorides}}.
\newblock {\emph{\JournalTitle{J. Chem. Phys.}}}
  \textbf{\bibinfo{volume}{155}}, \bibinfo{pages}{164103},
  \doiprefix\url{10.1063/5.0062927} (\bibinfo{year}{2021}).

\bibitem{Anderson}
\bibinfo{author}{Anderson, P.~W.}
\newblock \bibinfo{journal}{\bibinfo{title}{Localized {Magnetic} {States} in
  {Metals}}}.
\newblock {\emph{\JournalTitle{Phys. Rev.}}} \textbf{\bibinfo{volume}{124}},
  \bibinfo{pages}{41--53}, \doiprefix\url{10.1103/PhysRev.124.41}
  (\bibinfo{year}{1961}).

\bibitem{Wen}
\bibinfo{author}{Wen, X.-D.} \emph{et~al.}
\newblock \bibinfo{journal}{\bibinfo{title}{Effect of spin-orbit coupling on
  the actinide dioxides {AnO$_2$} ({An}={Th}, {Pa}, {U}, {Np}, {Pu}, and {Am}):
  {A} screened hybrid density functional study}}.
\newblock {\emph{\JournalTitle{The Journal of Chemical Physics}}}
  \textbf{\bibinfo{volume}{137}}, \bibinfo{pages}{154707},
  \doiprefix\url{10.1063/1.4757615} (\bibinfo{year}{2012}).

\bibitem{Suzuki}
\bibinfo{author}{Suzuki, M.-T.}, \bibinfo{author}{Magnani, N.} \&
  \bibinfo{author}{Oppeneer, P.~M.}
\newblock \bibinfo{journal}{\bibinfo{title}{Microscopic theory of the
  insulating electronic ground states of the actinide dioxides {AnO$_2$} ({An}
  = {U}, {Np}, {Pu}, {Am}, and {Cm})}}.
\newblock {\emph{\JournalTitle{Phys. Rev. B}}} \textbf{\bibinfo{volume}{88}},
  \bibinfo{pages}{195146}, \doiprefix\url{10.1103/PhysRevB.88.195146}
  (\bibinfo{year}{2013}).

\bibitem{C_Suzuki}
\bibinfo{author}{Suzuki, C.} \emph{et~al.}
\newblock \bibinfo{journal}{\bibinfo{title}{{DFT} study on the electronic
  structure and chemical state of {Americium} in an ({Am},{U}) mixed oxide}}.
\newblock {\emph{\JournalTitle{Journal of Physics and Chemistry of Solids}}}
  \textbf{\bibinfo{volume}{74}}, \bibinfo{pages}{1769--1774},
  \doiprefix\url{10.1016/j.jpcs.2013.07.006} (\bibinfo{year}{2013}).

\bibitem{Lu}
\bibinfo{author}{Lu, Y.}, \bibinfo{author}{Yang, Y.}, \bibinfo{author}{Zheng,
  F.}, \bibinfo{author}{Wang, B.-T.} \& \bibinfo{author}{Zhang, P.}
\newblock \bibinfo{journal}{\bibinfo{title}{Electronic, mechanical, and
  thermodynamic properties of americium dioxide}}.
\newblock {\emph{\JournalTitle{Journal of Nuclear Materials}}}
  \textbf{\bibinfo{volume}{441}}, \bibinfo{pages}{411--420},
  \doiprefix\url{10.1016/j.jnucmat.2013.06.043} (\bibinfo{year}{2013}).

\bibitem{Pegg}
\bibinfo{author}{Pegg, J.~T.}, \bibinfo{author}{Aparicio-Anglès, X.},
  \bibinfo{author}{Storr, M.} \& \bibinfo{author}{de~Leeuw, N.~H.}
\newblock \bibinfo{journal}{\bibinfo{title}{{DFT}+{U} study of the structures
  and properties of the actinide dioxides}}.
\newblock {\emph{\JournalTitle{Journal of Nuclear Materials}}}
  \textbf{\bibinfo{volume}{492}}, \bibinfo{pages}{269--278},
  \doiprefix\url{10.1016/j.jnucmat.2017.05.025} (\bibinfo{year}{2017}).

\bibitem{Noutack}
\bibinfo{author}{Talla~Noutack, M.~S.}, \bibinfo{author}{Geneste, G.},
  \bibinfo{author}{Jomard, G.} \& \bibinfo{author}{Freyss, M.}
\newblock \bibinfo{journal}{\bibinfo{title}{First-principles investigation of
  the bulk properties of americium dioxide and sesquioxides}}.
\newblock {\emph{\JournalTitle{Phys. Rev. Materials}}}
  \textbf{\bibinfo{volume}{3}}, \bibinfo{pages}{035001},
  \doiprefix\url{10.1103/PhysRevMaterials.3.035001} (\bibinfo{year}{2019}).

\bibitem{Chen}
\bibinfo{author}{Chen, J.-L.} \& \bibinfo{author}{Kaltsoyannis, N.}
\newblock \bibinfo{journal}{\bibinfo{title}{Computational {Study} of the {Bulk}
  and {Surface} {Properties} of {Minor} {Actinide} {Dioxides} {MAnO$_2$} ({MAn}
  = {Np}, {Am}, and {Cm}); {Water} {Adsorption} on {Stoichiometric} and
  {Reduced} \{111\}, \{110\}, and \{100\} {Surfaces}}}.
\newblock {\emph{\JournalTitle{J. Phys. Chem. C}}}
  \textbf{\bibinfo{volume}{123}}, \bibinfo{pages}{15540--15550},
  \doiprefix\url{10.1021/acs.jpcc.9b02324} (\bibinfo{year}{2019}).

\bibitem{Moree}
\bibinfo{author}{Morée, J.-B.}, \bibinfo{author}{Outerovitch, R.} \&
  \bibinfo{author}{Amadon, B.}
\newblock \bibinfo{journal}{\bibinfo{title}{First-principles calculation of the
  {Coulomb} interaction parameters {U} and {J} for actinide dioxides}}.
\newblock {\emph{\JournalTitle{Phys. Rev. B}}} \textbf{\bibinfo{volume}{103}},
  \bibinfo{pages}{045113}, \doiprefix\url{10.1103/PhysRevB.103.045113}
  (\bibinfo{year}{2021}).

\bibitem{Teterin}
\bibinfo{author}{Teterin, Y.} \emph{et~al.}
\newblock \bibinfo{journal}{\bibinfo{title}{X-ray photoelectron spectra
  structure and chemical bonding in {AmO$_2$}}}.
\newblock {\emph{\JournalTitle{Nucl Technol Radiat Prot}}}
  \textbf{\bibinfo{volume}{30}}, \bibinfo{pages}{83--98},
  \doiprefix\url{10.2298/NTRP1502083T} (\bibinfo{year}{2015}).

\bibitem{Moore}
\bibinfo{author}{Moore, K.~T.} \& \bibinfo{author}{van~der Laan, G.}
\newblock \bibinfo{journal}{\bibinfo{title}{Nature of the 5f states in actinide
  metals}}.
\newblock {\emph{\JournalTitle{Rev. Mod. Phys.}}}
  \textbf{\bibinfo{volume}{81}}, \bibinfo{pages}{235--298},
  \doiprefix\url{10.1103/RevModPhys.81.235} (\bibinfo{year}{2009}).

\bibitem{Moore02}
\bibinfo{author}{Moore, K.~T.}, \bibinfo{author}{van~der Laan, G.},
  \bibinfo{author}{Haire, R.~G.}, \bibinfo{author}{Wall, M.~A.} \&
  \bibinfo{author}{Schwartz, A.~J.}
\newblock \bibinfo{journal}{\bibinfo{title}{Oxidation and aging in {U} and {Pu}
  probed by spin-orbit sum rule analysis: {Indications} for covalent
  metal-oxide bonds}}.
\newblock {\emph{\JournalTitle{Phys. Rev. B}}} \textbf{\bibinfo{volume}{73}},
  \bibinfo{pages}{033109}, \doiprefix\url{10.1103/PhysRevB.73.033109}
  (\bibinfo{year}{2006}).

\bibitem{Butorin08}
\bibinfo{author}{Butorin, S.~M.} \emph{et~al.}
\newblock \bibinfo{journal}{\bibinfo{title}{X-ray spectroscopic study of
  chemical state in uranium carbides}}.
\newblock {\emph{\JournalTitle{J Synchrotron Rad}}}
  \textbf{\bibinfo{volume}{29}}, \bibinfo{pages}{295--302},
  \doiprefix\url{10.1107/S160057752101314X} (\bibinfo{year}{2022}).

\bibitem{Mayer}
\bibinfo{author}{Mayer, K.}, \bibinfo{author}{Kanellakopoulos, B.},
  \bibinfo{author}{Naegele, J.} \& \bibinfo{author}{Koch, L.}
\newblock \bibinfo{journal}{\bibinfo{title}{On the valency state of americium
  in (u$_{0.5}$am$_{0.5}$)o$_{2-x}$}}.
\newblock {\emph{\JournalTitle{Journal of Alloys and Compounds}}}
  \textbf{\bibinfo{volume}{213-214}}, \bibinfo{pages}{456--459},
  \doiprefix\url{10.1016/0925-8388(94)90960-1} (\bibinfo{year}{1994}).

\bibitem{Kvashnina03}
\bibinfo{author}{Kvashnina, K.~O.} \& \bibinfo{author}{Butorin, S.~M.}
\newblock \bibinfo{journal}{\bibinfo{title}{High-energy resolution {X}-ray
  spectroscopy at actinide {M$_{4,5}$} and ligand {K} edges: what we know, what
  we want to know, and what we can know}}.
\newblock {\emph{\JournalTitle{Chem. Commun.}}} \textbf{\bibinfo{volume}{58}},
  \bibinfo{pages}{327--342}, \doiprefix\url{10.1039/D1CC04851A}
  (\bibinfo{year}{2022}).

\bibitem{Yamazaki}
\bibinfo{author}{Yamazaki, T.} \& \bibinfo{author}{Kotani, A.}
\newblock \bibinfo{journal}{\bibinfo{title}{Systematic {Analysis} of 4
  \textit{f} {Core} {Photoemission} {Spectra} in {Actinide} {Oxides}}}.
\newblock {\emph{\JournalTitle{J. Phys. Soc. Jpn.}}}
  \textbf{\bibinfo{volume}{60}}, \bibinfo{pages}{49--52},
  \doiprefix\url{10.1143/JPSJ.60.49} (\bibinfo{year}{1991}).

\bibitem{Hubert}
\bibinfo{author}{Hubert, S.}, \bibinfo{author}{Thouvenot, P.} \&
  \bibinfo{author}{Edelstein, N.}
\newblock \bibinfo{journal}{\bibinfo{title}{Spectroscopic studies and
  crystal-field analyses of {Am$^{3+}$} and {Eu$^{3+}$} in the cubic-symmetry
  site of {ThO$_2$}}}.
\newblock {\emph{\JournalTitle{Phys. Rev. B}}} \textbf{\bibinfo{volume}{48}},
  \bibinfo{pages}{5751--5760}, \doiprefix\url{10.1103/PhysRevB.48.5751}
  (\bibinfo{year}{1993}).

\bibitem{Campbell}
\bibinfo{author}{Campbell, J.} \& \bibinfo{author}{Papp, T.}
\newblock \bibinfo{journal}{\bibinfo{title}{{WIDTHS} {OF} {THE} {ATOMIC}
  {K}–{N$_7$} {LEVELS}}}.
\newblock {\emph{\JournalTitle{Atomic Data and Nuclear Data Tables}}}
  \textbf{\bibinfo{volume}{77}}, \bibinfo{pages}{1--56},
  \doiprefix\url{10.1006/adnd.2000.0848} (\bibinfo{year}{2001}).

\bibitem{Ogasawara}
\bibinfo{author}{Ogasawara, H.}, \bibinfo{author}{Kotani, A.} \&
  \bibinfo{author}{Thole, B.~T.}
\newblock \bibinfo{journal}{\bibinfo{title}{Calculation of magnetic x-ray
  dichroism in \textit{4d} and \textit{5d} absorption spectra of actinides}}.
\newblock {\emph{\JournalTitle{Phys. Rev. B}}} \textbf{\bibinfo{volume}{44}},
  \bibinfo{pages}{2169--2181}, \doiprefix\url{10.1103/PhysRevB.44.2169}
  (\bibinfo{year}{1991}).

\bibitem{Kotani02}
\bibinfo{author}{Kotani, A.} \& \bibinfo{author}{Ogasawara, H.}
\newblock \bibinfo{journal}{\bibinfo{title}{Theory of core-level spectroscopy
  in actinide systems}}.
\newblock {\emph{\JournalTitle{Physica B: Condensed Matter}}}
  \textbf{\bibinfo{volume}{186-188}}, \bibinfo{pages}{16--20},
  \doiprefix\url{10.1016/0921-4526(93)90485-O} (\bibinfo{year}{1993}).

\bibitem{Butorin05}
\bibinfo{author}{Butorin, S.~M.}
\newblock \bibinfo{title}{Resonant {Inelastic} {Soft} {X}-{Ray} {Scattering}
  {Spectroscopy} of {Light}-{Actinide} {Materials}}.
\newblock In \bibinfo{editor}{Kalmykov, S.~N.} \& \bibinfo{editor}{Denecke,
  M.~A.} (eds.) \emph{\bibinfo{booktitle}{Actinide {Nanoparticle} {Research}}},
  \bibinfo{pages}{63--103}, \doiprefix\url{10.1007/978-3-642-11432-8_3}
  (\bibinfo{publisher}{Springer Berlin Heidelberg}, \bibinfo{address}{Berlin,
  Heidelberg}, \bibinfo{year}{2011}).

\bibitem{Veal}
\bibinfo{author}{Veal, B.~W.}, \bibinfo{author}{Lam, D.~J.},
  \bibinfo{author}{Diamond, H.} \& \bibinfo{author}{Hoekstra, H.~R.}
\newblock \bibinfo{journal}{\bibinfo{title}{X-ray photoelectron-spectroscopy
  study of oxides of the transuranium elements {Np}, {Pu}, {Am}, {Cm}, {Bk},
  and {Cf}}}.
\newblock {\emph{\JournalTitle{Phys. Rev. B}}} \textbf{\bibinfo{volume}{15}},
  \bibinfo{pages}{2929--2942}, \doiprefix\url{10.1103/PhysRevB.15.2929}
  (\bibinfo{year}{1977}).

\bibitem{Nevitt}
\bibinfo{author}{Nevitt, P.}
\newblock \emph{\bibinfo{title}{Photoemission studies of the light actinides}}.
\newblock Ph.D. thesis, \bibinfo{school}{Cardiff University}
  (\bibinfo{year}{2005}).

\bibitem{Gouder}
\bibinfo{author}{Gouder, T.}, \bibinfo{author}{Oppeneer, P.~M.},
  \bibinfo{author}{Huber, F.}, \bibinfo{author}{Wastin, F.} \&
  \bibinfo{author}{Rebizant, J.}
\newblock \bibinfo{journal}{\bibinfo{title}{Photoemission study of the
  electronic structure of {Am}, {AmN}, {AmSb}, and {Am$_2$O$_3$} films}}.
\newblock {\emph{\JournalTitle{Phys. Rev. B}}} \textbf{\bibinfo{volume}{72}},
  \bibinfo{pages}{115122}, \doiprefix\url{10.1103/PhysRevB.72.115122}
  (\bibinfo{year}{2005}).

\bibitem{Finck}
\bibinfo{author}{Finck, N.} \emph{et~al.}
\newblock \bibinfo{journal}{\bibinfo{title}{{XAS} signatures of {Am({III})}
  adsorbed onto magnetite and maghemite}}.
\newblock {\emph{\JournalTitle{J. Phys.: Conf. Ser.}}}
  \textbf{\bibinfo{volume}{712}}, \bibinfo{pages}{012085},
  \doiprefix\url{10.1088/1742-6596/712/1/012085} (\bibinfo{year}{2016}).

\bibitem{Zaanen}
\bibinfo{author}{Zaanen, J.}, \bibinfo{author}{Sawatzky, G.~A.} \&
  \bibinfo{author}{Allen, J.~W.}
\newblock \bibinfo{journal}{\bibinfo{title}{Band gaps and electronic structure
  of transition-metal compounds}}.
\newblock {\emph{\JournalTitle{Phys. Rev. Lett.}}}
  \textbf{\bibinfo{volume}{55}}, \bibinfo{pages}{418--421},
  \doiprefix\url{10.1103/PhysRevLett.55.418} (\bibinfo{year}{1985}).

\bibitem{Moody}
\bibinfo{author}{Moody, K.~J.} \emph{et~al.}
\newblock \bibinfo{title}{Analytical {Chemistry} of {Plutonium}}.
\newblock In \bibinfo{editor}{Morss, L.~R.}, \bibinfo{editor}{Edelstein, N.~M.}
  \& \bibinfo{editor}{Fuger, J.} (eds.) \emph{\bibinfo{booktitle}{The
  {Chemistry} of the {Actinide} and {Transactinide} {Elements}}},
  \bibinfo{pages}{3889--4003}, \doiprefix\url{10.1007/978-94-007-0211-0_36}
  (\bibinfo{publisher}{Springer Netherlands}, \bibinfo{address}{Dordrecht},
  \bibinfo{year}{2010}).

\bibitem{Kvashnina_Cm}
\bibinfo{author}{Kvashnina, K.~O.} \emph{et~al.}
\newblock \bibinfo{journal}{\bibinfo{title}{Resonant inelastic x-ray scattering
  of curium oxide}}.
\newblock {\emph{\JournalTitle{Phys. Rev. B}}} \textbf{\bibinfo{volume}{75}},
  \bibinfo{pages}{115107}, \doiprefix\url{10.1103/PhysRevB.75.115107}
  (\bibinfo{year}{2007}).

\bibitem{Shuh}
\bibinfo{author}{Smiles, D.~E.} \& \bibinfo{author}{Shuh, D.~K.}
\newblock \bibinfo{title}{Soft x-ray synchrotron radiation studies of plutonium
  materials}.
\newblock In \bibinfo{editor}{Clark, D.~L.}, \bibinfo{editor}{Geeson, D.~A.} \&
  \bibinfo{editor}{Hanrahan, J., R.~J.} (eds.)
  \emph{\bibinfo{booktitle}{Plutonium Handbook, 2nd Edition}},
  vol.~\bibinfo{volume}{6}, \bibinfo{pages}{2991--3007}
  (\bibinfo{publisher}{American Nuclear Society}, \bibinfo{year}{2019}).

\bibitem{Modin}
\bibinfo{author}{Modin, A.} \emph{et~al.}
\newblock \bibinfo{journal}{\bibinfo{title}{Closed source experimental system
  for soft x-ray spectroscopy of radioactive materials}}.
\newblock {\emph{\JournalTitle{Rev. Sci. Instrum.}}}
  \textbf{\bibinfo{volume}{79}}, \bibinfo{pages}{093103},
  \doiprefix\url{10.1063/1.2991109} (\bibinfo{year}{2008}).

\bibitem{Denecke}
\bibinfo{author}{Denecke, R.} \emph{et~al.}
\newblock \bibinfo{journal}{\bibinfo{title}{Beamline {I511} at {MAX} {II},
  capabilities and performance}}.
\newblock {\emph{\JournalTitle{Journal of Electron Spectroscopy and Related
  Phenomena}}} \textbf{\bibinfo{volume}{101-103}}, \bibinfo{pages}{971--977},
  \doiprefix\url{10.1016/S0368-2048(98)00358-2} (\bibinfo{year}{1999}).

\bibitem{Warwick}
\bibinfo{author}{Warwick, T.}, \bibinfo{author}{Heimann, P.},
  \bibinfo{author}{Mossessian, D.}, \bibinfo{author}{McKinney, W.} \&
  \bibinfo{author}{Padmore, H.}
\newblock \bibinfo{journal}{\bibinfo{title}{Performance of a high resolution,
  high flux density {SGM} undulator beamline at the {ALS} (invited)}}.
\newblock {\emph{\JournalTitle{Review of Scientific Instruments}}}
  \textbf{\bibinfo{volume}{66}}, \bibinfo{pages}{2037--2040},
  \doiprefix\url{10.1063/1.1145789} (\bibinfo{year}{1995}).

\bibitem{Kotani}
\bibinfo{author}{Kotani, A.} \& \bibinfo{author}{Ogasawara, H.}
\newblock \bibinfo{journal}{\bibinfo{title}{Theory of core-level spectroscopy
  of rare-earth oxides}}.
\newblock {\emph{\JournalTitle{Journal of Electron Spectroscopy and Related
  Phenomena}}} \textbf{\bibinfo{volume}{60}}, \bibinfo{pages}{257--299},
  \doiprefix\url{10.1016/0368-2048(92)80024-3} (\bibinfo{year}{1992}).

\bibitem{Butorin06}
\bibinfo{author}{Butorin, S.~M.} \emph{et~al.}
\newblock \bibinfo{journal}{\bibinfo{title}{Resonant {X}-{Ray} {Fluorescence}
  {Spectroscopy} of {Correlated} {Systems}: {A} {Probe} of {Charge}-{Transfer}
  {Excitations}}}.
\newblock {\emph{\JournalTitle{Phys. Rev. Lett.}}}
  \textbf{\bibinfo{volume}{77}}, \bibinfo{pages}{574--577},
  \doiprefix\url{10.1103/PhysRevLett.77.574} (\bibinfo{year}{1996}).

\bibitem{Nakazawa}
\bibinfo{author}{Nakazawa, M.}, \bibinfo{author}{Ogasawara, H.} \&
  \bibinfo{author}{Kotani, A.}
\newblock \bibinfo{journal}{\bibinfo{title}{{THEORY} {OF} {POLARIZATION}
  {DEPENDENCE} {IN} {RESONANT} {X}-{RAY} {EMISSION} {SPECTRA} {OF} {A}
  {URANIUM} {COMPOUND}}}.
\newblock {\emph{\JournalTitle{Surf. Rev. Lett.}}}
  \textbf{\bibinfo{volume}{09}}, \bibinfo{pages}{1149--1153},
  \doiprefix\url{10.1142/S0218625X02003433} (\bibinfo{year}{2002}).

\bibitem{Butorin07}
\bibinfo{author}{Butorin, S.~M.}, \bibinfo{author}{Modin, A.},
  \bibinfo{author}{Vegelius, J.~R.}, \bibinfo{author}{Kvashnina, K.~O.} \&
  \bibinfo{author}{Shuh, D.~K.}
\newblock \bibinfo{journal}{\bibinfo{title}{Probing {Chemical} {Bonding} in
  {Uranium} {Dioxide} by {Means} of {High}-{Resolution} {X}-ray {Absorption}
  {Spectroscopy}}}.
\newblock {\emph{\JournalTitle{J. Phys. Chem. C}}}
  \textbf{\bibinfo{volume}{120}}, \bibinfo{pages}{29397--29404},
  \doiprefix\url{10.1021/acs.jpcc.6b09335} (\bibinfo{year}{2016}).

\bibitem{Cowan}
\bibinfo{author}{Cowan, R.~D.}
\newblock \emph{\bibinfo{title}{The theory of atomic structure and spectra}}.
\newblock No.~\bibinfo{number}{3} in \bibinfo{series}{Los {Alamos} series in
  basic and applied sciences} (\bibinfo{publisher}{University of California
  Press}, \bibinfo{address}{Berkeley}, \bibinfo{year}{1981}).

\bibitem{Butler}
\bibinfo{author}{Butler, P.~H.}
\newblock \emph{\bibinfo{title}{Point {Group} {Symmetry} {Applications}:
  {Methods} and {Tables}.}} (\bibinfo{publisher}{Springer US},
  \bibinfo{address}{Boston}, \bibinfo{year}{1981}).
\newblock \bibinfo{note}{OCLC: 958528839}.

\bibitem{Thole}
\bibinfo{author}{Thole, B.}, \bibinfo{author}{Van Der~Laan, G.} \&
  \bibinfo{author}{Butler, P.}
\newblock \bibinfo{journal}{\bibinfo{title}{Spin-mixed ground state of {Fe}
  phthalocyanine and the temperature-dependent branching ratio in {X}-ray
  absorption spectroscopy}}.
\newblock {\emph{\JournalTitle{Chemical Physics Letters}}}
  \textbf{\bibinfo{volume}{149}}, \bibinfo{pages}{295--299},
  \doiprefix\url{10.1016/0009-2614(88)85029-2} (\bibinfo{year}{1988}).

\end{thebibliography}

\end{document}